\documentclass{PoS}

\title{Constraining dark matter signal from a combined analysis of Milky Way satellites using the Fermi-LAT }

\ShortTitle{dSph stacking with the Fermi-LAT }

\author{\speaker{Maja Llena Garde}%
        \\
        Stockholm University\\
        E-mail: \email{maja.garde@fysik.su.se}}
\author{On behalf of the Fermi-LAT collaboration}

\abstract{The Fermi LAT collaboration has recently presented constraints on the gamma-ray signal from annihilating dark matter using separate analyses of a number of dwarf spheroidal galaxies. Since the expected annihilation signal has the same physical properties regardless of the target (except for a normalization scale), it is possible to enhance the constraining power using a combined analysis, the initial results of which will be presented here.}

\FullConference{Identification of Dark Matter 2010\\
		 July 26 - 30 2010\\
		 University of Montpellier 2, Montpellier, France}

\begin{document}



\section{Introduction}
The Fermi Gamma-ray Space Telescope was launched on June 11, 2008. Its main instrument, the Large Area Telescope (Fermi-LAT), consists of 16 identical modules in a 4x4 array with each module comprised of a tracker for direction determination and a calorimeter for energy measurements \cite{Atwood:2009ez}. The Fermi-LAT observes the entire sky every $\sim$3 hours (2 orbits) with a field of view covering $\sim$2.4 sr and a sensitive energy range extending from 20MeV to >300GeV. These features makes the Fermi-LAT a great instrument for dark matter (DM) searches. 

The leading DM candidate is a weakly interacting massive particle (WIMP), and this is the DM candidate we are primarily focusing on with the Fermi-LAT. The gamma-ray flux from self-annihilating WIMPs can be expressed as $\phi_{WIMP} (E, \psi) = J(\psi)\times\Phi^{PP}(E)$ \cite{Baltz:2008wd}, where $\Phi^{PP}(E)$ is the "particle physics factor" described by
\begin{equation}
\Phi^{PP}(E)=\frac{1}{2}\frac{<\sigma v>}{4\pi m_{WIMP}^{2}}\sum _f \frac{dN_f}{dE}B_f
\end{equation}
and $J(\psi)$ is the "astrophysical factor", or J-factor, described by
\begin{equation}
J(\psi)=\int_{l.o.s.}dl(\psi)\rho^{2}(l(\psi)).
\end{equation}
Here $\left\langle \sigma  v \right\rangle$ is the WIMP annihilation cross section times relative velocity, $m_{WIMP}$ is the WIMP mass, $\rho(r)$ is the dark matter density distribution, and $\sum _f \frac{dN_f}{dE}B_f$ is the gamma ray spectrum generated per WIMP annihilation where the sum is over final states $f$ with branching ratio $B_f$. The particle physics factor has two spectral features: the continuum feature and the line feature (which is often referred to as the ''smoking gun'').

Dwarf spheroidal galaxies (dSphs) are DM dominated systems since they have a high mass to light ratio. Many dSphs are closer than 100 kpc to the galactic center, and they have low background since most dSphs are expected to be free from other astrophysical gamma-ray sources and they have a small gas content. This makes them a good target for gamma-ray DM searches with the Fermi-LAT. However, the photon statistics for single dSphs are expected to be very low. The Fermi-LAT collaboration has recently presented results from a DM search in a number of dSphs \cite{Abdo:2010ex}. Since the DM spectra are the same for all astrophysical sources, a combined analysis would improve the statistics.

In this proceeding we present the preliminary results of a combined likelihood analysis of eight dSphs.

\section{Analysis}

\begin{table}
\begin{center}
\begin{tabular}{llll}
Name & l &  b & $J^{NFW} (\times10^{19} \frac{GeV^2}{cm^5})$ \\ \hline
Bootes I & $358.08$ & $69.62$ & $0.16^{+0.35}_{-0.13}$ \\
Coma Berenices & $241.9$ & $83.6$ & $0.16^{+0.22}_{-0.08}$ \\
Draco & $86.37$ & $34.72$ & $1.20^{+0.31}_{-0.25}$ \\
Fornax & $237.1$ & $-65.7$ & $0.06^{+0.03}_{-0.03}$ \\
Sculptor & $287.15$ & $-83.16$ & $0.24^{+0.06}_{-0.06}$ \\
Sextans & $243.4$ & $42.2$ & $0.06^{+0.03}_{-0.02}$ \\
Ursa Mayor II & $152.46$ & $37.44$ & $0.58^{+0.91}_{-0.35}$ \\
Ursa Minor & $104.95$ & $44.80$ & $0.64^{+0.25}_{-0.18}$ \\
\end{tabular}
\caption{dSphs analyzed in this work. All positions and J-factors of the dSphs are taken from \cite{Abdo:2010ex}. \label{tab1}}
\end{center}
\end{table}
The combined likelihood used in this work is described by
\begin{equation}
 L(\sigma v,m_{WIMP}; obs)=\displaystyle\prod_{i}^{N} L_{i}(\sigma v,m_{WIMP}, C, b_i; obs),
\end{equation}
where $\sigma v$ and $m_{WIMP}$ are the common DM signal parameters (velocity averaged cross-section and mass), $C$ are constants (e.g. branching fraction in our case), and $b$ are individual parameters (isotropic and galactic diffuse background normalization, normalization of nearby sources). The main advantages of the combined likelihood are that the analysis can be individually optimized and that combined limits are more robust under individual background fluctuations and under individual astrophysical modelling uncertainties than individual limits.

We have observed the eight dSphs listed in Table \ref{tab1}, which is the same subset analyzed in \cite{Abdo:2010ex}. In this preliminary analysis we have used 21 month data from 2008-08-04 to 2010-05-12. We have used the diffuse event class which only contains the events with the highest gamma-like confidence, and we have chosen events ranging from 200MeV to 100GeV. We used the Fermi-LAT instrument responce function P6$\_$V3$\_$DIFFUSE. Our region of interest (ROI) is a region of 10 degrees radius centered on dSph location. Standard cuts removing Earth albedo photons have been made. The dSphs are modeled as DM point sources using the DMFit package \cite{Jeltema:2008hf} where we only consider 100\% annihilation into the b\={b} annihilation channel so far since it represents the spectral shape of a large fraction of the possible final states. The background is modeled according to Fermi-LAT recommendations \cite{modelPage}, and nearby soures are modeled according to the first year point source catalog \cite{Collaboration:2010ru}. We perform a binned analysis to use both energy and spatial information. The data selection and analysis were performed using the Fermi-LAT analysis package, ScienceTools \cite{STpage}, and the upper limits are obtained using profile likelihood as implemented in the MINUIT processor MINOS \cite{minuit}.

In this proceeding work we have not yet taken systematic uncertainties into account. The uncertainties mainly arise from the J-factors (about 20 to 200\% as can be seen in Table \ref{tab1}), and the effective area of the LAT (5 to 20\% and energy dependent \cite{LATcaveats}). However, some initial tests show that the combined results are less sensitive to J-factor changes than individual limits.

\section{Preliminary results}
\begin{figure}
\begin{center}
\includegraphics[width=.6\textwidth]{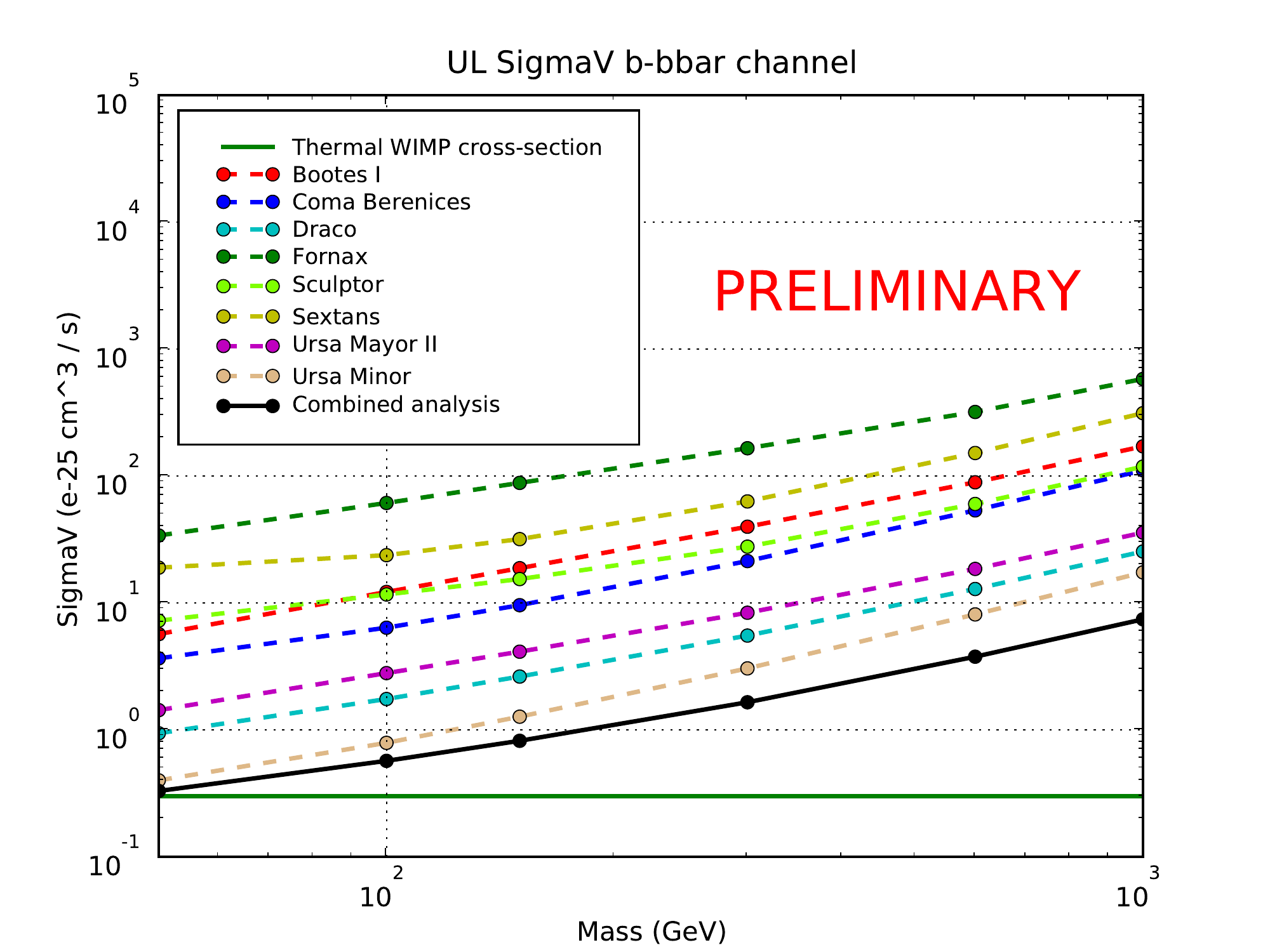}
\caption{Upper limits on WIMP annihilation cross section for annihilation into 100\% b\={b}, $\sigma v$ evaluated at $m_{WIMP}$ = 50, 100, 150, 300, 600 and 1000 GeV. The expected thermal WIMP cross-section is plotted as a reference. The limits are 84\% two-sided. \label{fig1}}
\end{center}
\end{figure}
In Fig. \ref{fig1} we present constraints on the WIMP annihilation cross-section for eight dSphs, both individual and combined limits, evaluated at $m_{WIMP}$ = 50, 100, 150, 300, 600 and 1000 GeV. No systematic uncertainties have been included and the results are preliminary. The combined limits approach the parameter space below the thermal WIMP cross-section for low WIMP masses. Limits down to 5 GeV were presented at the conference, but due to an extrapolation problem for masses below 10 GeV only limits for mass points above 10 GeV are reliable. An update for low masses is forthcoming.

\section{Discussion and outlook}
A stacking analysis of 8 dSphs has been presented using a combined likelihood approach. Limits improve with respect to the the most stringent of the 8 individual limits, depending on WIMP mass. Some tests on consistency under choice of ROI, fit range and binning have been performed. The results presented here are preliminary. The inclusion of systematic uncertainties is in progress. More dSphs will be added if they are sufficiently nearby, if relatively accurate estimates of their DM distribution can be obtained, and if they are situated at sufficiently high galactic latitude to avoid galactic foreground. The analysis will be updated for more recent data, more annihilation channels will be studied, and a paper is in preparation within the Fermi-LAT collaboration.

\begin{acknowledgments}
The Fermi LAT Collaboration acknowledges generous ongoing support from a number of agencies and institutes that have supported both the development and the operation of the LAT as well as scientific data analysis. These include the National Aeronautics and Space Administration and the Department of Energy in the United States, the Commissariat \'a l'Energie Atomique and the Centre National de la Recherche Scientifique / Institut National de Physique Nucl\'eaire et de Physique des Particules in France, the Agenzia Spaziale Italiana and the Istituto Nazionale di Fisica Nucleare in Italy, the Ministry of Education, Culture, Sports, Science and Technology (MEXT), High Energy Accelerator Research Organization (KEK) and Japan Aerospace Exploration Agency (JAXA) in Japan, and the K. A. Wallenberg Foundation, the Swedish Research Council and the Swedish National Space Board in Sweden.
\end{acknowledgments}

\end{document}